\documentclass[11pt]{article}
\usepackage{moriond,epsfig}

\def\be{\begin{equation}}
\def\ee{\end{equation}}
\def\bea{\begin{eqnarray}}
\def\eea{\end{eqnarray}}

\begin{document}
\vspace*{4cm}
\title{Comment about the "Gravity coupled to a scalar field in extra dimensions"
paper.}

\author{S.M.KOZYREV}

\address{Scientific center gravity wave studies ''Dulkyn''.}

\maketitle\abstracts{Wehus and Ravndal~\cite{ligo} have argued that in d + 1 dimensions
the general solution for gravity minimally coupled to a scalar field can not
be explicitly written in Schwarzschild coordinates. We contest these objections.}

There has been a revival of interest in scalar fields theory in recent times, particularly
 in the context of string theory. The continuing focus on extra-dimensional models in
fundamental physics provides motivation for studying scalar fields in higher dimensions.
The physical relevance of scalar fields in todays gravitational physics and cosmology also
stems from their role in current cosmological models~\cite{geo}. It is important to note in
this connection that the widely employed equivalence between (D+1) Kaluza-Klein theories
with empty D-dimensional Jordan Brans-Dicke theories $({\omega} = 0)$ lead us to
considerations a scalar field. Thus in this sense dimensional reduction ''generates''
sources which in lower dimension and can be interpreted as scalar field.

In a recent paper, Wehus and Ravndal~\cite{ligo}, provide a good overview of
the various Einstein-scalar field solutions and suggested that the general
solution of the equations for gravity coupled minimally to a massless scalar
field can not be explicitly written in Schwarzschild coordinates. The
purpose of this short comment is to demonstrate that these solutions can be
alternatively derived by exploiting the change of variables technique. First
time, this technique was applied to a spherically symmetric case for finding
a new solution to the Jordan Brans-Dicke-scalar field equations~\cite{virgo}.
Techniques for obtaining the similar static solutions are known by know~\cite{bursts4}.
In D = d + 1 dimensions, the action for a scalar field minimally
coupled to gravity given by (we take units G = c = 1):
\begin{equation}
S=\int d^D\sqrt{-g}\frac 12\left( R-g^{\mu \mu }\phi _{,\mu }\phi _{,\nu
}\right) .
\label{e1}
\end{equation}

One can get the minimally coupled Einstein-scalar fields equations by
variation the above action:
\begin{equation}
R_{\mu \nu }-\frac 12Rg_{\mu \nu }=\left( \nabla _\mu \phi \nabla _\nu \phi
-\frac 12g_{\mu \nu }g^{\alpha \beta }\nabla _\alpha \phi \nabla _\beta \phi
\right) ,  \label{e2}
\end{equation}

\begin{equation}
\phi =0.  \label{e3}
\end{equation}

$\nabla _\alpha $ is the covariant derivative associated with the metric g,
R$_{\mu \mu }$ and R are the Ricci tensor and Rieci scalar for an arbitrary metric g.

By contracting equation (\ref{e1}), we can rewrite this equation as
\begin{equation}
R_{\mu \nu }=\nabla _\mu \phi \nabla _\nu \phi ,  \label{e4}
\end{equation}

As we have already mentioned we consider standard static and spherically
symmetric space-time. The static and spherically symmetric metric in
Schwarzschild coordinates for d + 1 dimensions can be written as

\begin{equation}
ds^2=-e^{2\alpha (r)}dt^2+e^{2\beta (r)}dr^2+r^2d\Omega _d^2.   \label{e5}
\end{equation}

where $\alpha $, $\beta $ are unknown functions of the radial coordinate r,
and d$\Omega _d^2$ is the solid angle element in d - 1 dimensions. In order
to simplify the problem of solving the field equations we will replace
variable {\it r }by {\it r}($\alpha $) then the field equations (\ref{e4})
take a form:

\begin{equation}
1-\left( 1+\left( d-1\right) \frac{r^{\prime }\left( \alpha \right) }{%
r\left( \alpha \right) }\right) \lambda ^{\prime }\left( \alpha \right) -%
\frac{r^{"}\left( \alpha \right) }{r^{\prime }\left( \alpha \right) }=-\phi
^{\prime }\left( \alpha \right) ^2,  \label{e6}
\end{equation}

\begin{equation}
-1+\left( d-2\right) \left( e^{2\lambda \left( \alpha \right) }-1\right)
\frac{r^{\prime }\left( \alpha \right) }{r\left( \alpha \right) }+\lambda
^{\prime }\left( \alpha \right) =0,  \label{e7}
\end{equation}

\begin{equation}
-1+\left( d-1\right) \frac{r^{\prime }\left( \alpha \right) }{r\left( \alpha
\right) }-\frac{r^{"}\left( \alpha \right) }{r^{\prime }\left( \alpha
\right) }-\lambda ^{\prime }\left( \alpha \right) =0,  \label{e8}
\end{equation}

and the equation of motion for the scalar field (\ref{e3})

\begin{equation}
-1+\left( d-1\right) \frac{r^{\prime }\left( \alpha \right) }{r\left( \alpha
\right) }-\frac{r^{"}\left( \alpha \right) }{r^{\prime }\left( \alpha
\right) }-\lambda ^{\prime }\left( \alpha \right) +\frac{\phi ^{"}\left(
\alpha \right) }{\phi ^{\prime }\left( \alpha \right) }=0,  \label{e9}
\end{equation}

where $\alpha $ is a new variable and the primes denote derivatives with
respect to $\alpha $. Making use of equation (\ref{e8}) in  (\ref{e9}) they
simplify to

\begin{equation}
\frac{\phi ^{"}\left( \alpha \right) }{\phi ^{\prime }\left( \alpha \right) }%
=0,  \label{e10}
\end{equation}

thus we obtain

\begin{equation}
\phi \left( \alpha \right) =\zeta +\xi \alpha ,  \label{e11}
\end{equation}

where $\zeta $ and $\xi $ is a arbitrary constants. Using the asymptotic
condition in infinity we have $\zeta $ =1. In the case ${\it \xi }$ = 0 one
can find solution of equations  (\ref{e6}) - (\ref{e8})

\begin{equation}
r\left( \alpha \right) =\frac{const}{e^{2\alpha }-1},\ e^{\lambda \left(
\alpha \right) }=e^{-\alpha },\ \phi \left( \alpha \right) =1,  \label{e12}
\end{equation}

that identical to the Schwarzschild solution of the Einstein theory. For the
more general case $\xi \neq 0$ making use equations  (\ref{e6}) , (\ref{e8})
and (\ref{e11}) we eliminate $\lambda ^{\prime }\left( \alpha \right) $ and
obtain for {\it r}$\left( \alpha \right) $

\begin{center}
\begin{equation}
r\left( \alpha \right) =e^{-\frac \alpha {d-2}}\chi \ \cosh \left( \sqrt{1+%
\frac{\left( d-2\right) \xi ^2}{d-1}}\left( \alpha +\psi \right) \right)
^{\frac 1{d-2}},  \label{e13}
\end{equation}
\end{center}

where $\chi $ and $\psi $ is a arbitrary constants. After same algebra one
can find from (\ref{e7}) - (\ref{e8}) for $\lambda \left( \alpha \right) :$

\begin{equation}
\begin{array}{l}
\lambda \left( \alpha \right) =-\frac 12\log \left( d-2\right) -\log \left(
r^{\prime }\left( \alpha \right) \right) - \\
\,-\frac 12\log \left(
r\left( \alpha \right) +\left( d-1\right) r^{\prime }\left( \alpha \right)
\right) + \\
+\frac 12\log [\left( 2d-3\right) r\left( \alpha \right) r^{\prime }\left(
\alpha \right) +\left( d-2\right) \left( d-1\right) r^{\prime }\left( \alpha
\right) ^3+ \\
+r\left( \alpha \right) ^2r"\left( \alpha \right) -r^{\prime }\left( \alpha
\right) \xi ^2],
\end{array}
\label{e14}
\end{equation}

In the search for static and spherical symmetric solutions, we choose the
metric in Schwarzschild coordinates. This gives us both Schwarzschild
solution and more general solution of the minimally coupled equations. A
point to be noted is that in the Jordan, Brans-Dicke theory this change of
variables technique easily give us the arbitrary dimension Heckmann
solution, too.

\end{document}